\documentstyle[preprint,tighten,pra,aps,epsfig]{revtex}
\begin{document} \draft


\title{\LARGE \bf 
Planetary g(t) for which resistive atmospheric falling is rising }

\author{Haret C. Rosu}
\address{ Instituto de F\'{\i}sica,
Universidad de Guanajuato, Apdo Postal E-143, 37150 Le\'on, Gto, Mexico}

\author{Fermin Aceves de la Cruz}
\address{ Instituto de F\'{\i}sica,
Universidad de Guanajuato, Apdo Postal E-143, 37150 Le\'on, Gto, Mexico}

\maketitle

\begin{abstract}

\noindent 
A Darboux-transformed surface gravitational acceleration of the 
constant gravitational acceleration for a body
endowed with an atmospheric layer
is shown to turn the atmospheric free fall with quadratic resistance 
in the opposite motion, i.e., a free rising. Although the atmosphere
of such a body may look completely normal, it is the time dependence of
its gravitational field that produces this type of motion. This result
is a consequence of general, one-parameter-dependent Darboux transformations
in mathematical physics.

\end{abstract}


\section{Introduction}

\noindent
The following time-dependent gravitational acceleration
$$
g(t;\lambda)=g\Bigg[1-2\frac{d^2}{dt^2}
\ln(I_{01}(t) +\lambda)\Bigg]~,
\eqno(1)
$$
where $I_{01}(t)=\int _{0}^{t}{\rm cosh}^2xdx$ and $\lambda$ is a parameter of
the gravitational force, turns falling through a quadratic resistive 
- earth-like - atmosphere into just the opposite motion, i.e., a rising that
for $t\rightarrow \infty$ is of constant velocity.
Eq.~(1) is in fact the one-parameter Darboux-transformed acceleration of the constant 
acceleration $g$ at the surface of a usual planet such as Earth. 
The result is a consequence of a mathematical scheme that has been called 
strictly isospectral Darboux technique (SIDT) and has been applied extensively 
by one of the authors to various fields of physics.\cite{1} 
Briefly, SIDT is a three step procedure that for the resistive 
atmospheric free fall means the following.

(i) One starts with the equation 

$$
\frac{dv}{dt}=g-\epsilon v^2~,
\eqno(2)
$$ 
representing the 
normal resistive falling motion through an atmosphere of constant friction 
coefficient per unit of mass $\epsilon$. 
  
(ii)
One shifts to a motion of the form
$$
\frac{-dv}{dt}=g_2(t)- \epsilon v^2~, 
\eqno(3)
$$
where
$$
g_2(t)=g[-1+2\tanh ^2(\sqrt{\epsilon g}t)]~.
\eqno(4)
$$

One interpretation of Eq.~(3) is that the motion is upwards  
through a fluid medium producing a  driving
force proportional to the square of the velocity of the particle
(one may call it {\em antifriction}) plus another force $g_2(t)$ of driving 
character for $t<T=\frac{1}{\sqrt{\epsilon g}}{\rm Arctanh} 
\frac{1}{\sqrt{2}}$
and turning friction-like afterwards.

The first two steps are connected to each other through the particular 
solution of the initial normal motion as given by Eq.~(7) below. 

(iii)
The third step is the return back 
to an equation similar to the initial one, namely
$$
\frac{dv}{dt}=g(t; \lambda)- \epsilon v^2~, 
\eqno(5)
$$
where the time-dependent forces $g(t;\lambda)$ are given by
Eq.~(1).
The second and the third steps are also connected to each other through the 
general solution of the second problem (see Eq.~(16) below). 
Thus, although the motion given
by Eq.~(5) looks normal, its solution $v(t)$ corresponds to 
a rising on the vertical rather than a fall. Exactly as in the falling case,
the rising has a short accelerating stage followed by a uniform rising with
constant limiting velocity. Indeed, for sufficiently large times,
the driving force is 
at its maximum value $-\epsilon V^2$, whereas $g_2(t\rightarrow \infty)
=+g$ is a constant friction force and one gets the value of $V$ from
the balance of the two forces.

Here, we first review the usual resistive atmospheric falling according
to a 40-year old discussion of Davis.\cite{D} Next, we briefly describe
the motion in a Darboux-transformed gravitational field. Some concluding
remarks end up the work.

\bigskip

\section{ Davis' resistive free fall} 

\bigskip

\noindent
The free fall in quadratic resistive media has been studied by Davies in his
book as a simple application of the Riccati equation for the falling velocity. 
Davis wrote the  ``readily found" solution of Eq.~(2) in the following form
$$
v(t)=V\left(\frac{u_0+{\rm tanh} \sqrt{\epsilon g}t}{1+u_0{\rm tanh}
\sqrt{\alpha g} t}\right)~,
\eqno(6)
$$
where $u_0=v_0/V$ and $V=\sqrt{g/\epsilon}$ is the so-called limiting (terminal)
velocity.
Notice that Eq.~(6) can also be written as
$$
v(t)=V{\rm tanh}
(\sqrt{\epsilon g} t+ \beta)~,
\eqno(7)
$$
where the arbitrary phase $\beta$ is fixed through the initial condition,
$\beta ={\rm Arctanh} u_0$.

Davis obtained the falling distance $s_D(t)$ by integrating the velocity:
$$
s_D(t)=\int _{0}^{t}v(t)dt
=\frac{V^2}{g}\ln \left(\cosh \frac{gt}{V}+\frac{v_0}{V}\sinh \frac{gt}{V}
\right)~.
\eqno(8)
$$
The comparison with a set of 25 experimental data of parachute falls in the 
atmosphere of the earth was presented by Davis for the particular 
case $u_0=0$, i.e.
$
v_p=V\tanh\sqrt{\epsilon g}t
$
and
$$
s_D=\frac{V^2}{g}\ln \left(\cosh \frac{gt}{V}\right)
~\approx Vt-\frac{V^2}{g}\ln 2~,
\eqno(9)
$$
where the last approximation is for large times. The motion $s_D$ is 
asymptotically a uniform falling with the limiting velocity $V$.

\bigskip

\section{One-parameter Darboux transformed forces and resulting motion}

\bigskip

\noindent
We now provide details of the Darboux constructions.
We recall that 
nonrelativistic supersymmetric quantum mechanics 
is a simple 
application of Darboux transformations, 
with a huge publication output during the last two decades. One 
usually starts with a Riccati equation, such as Eq.~(2), that we write
in dimensionless form 
$$
{\cal R}_1: \quad \frac{du}{d\tau}+u^2=1~, 
\eqno(10)
$$
having $u_1=\tanh(\tau +\gamma)$ as solution.
Next, it is easy to see that the right hand side
corresponds up to a sign to a constant potential function for  
the Schr\"odinger equation at zero energy
$$
\left(\frac{d}{d\tau}+u_1\right)
\left(\frac{d}{d\tau}-u_1\right)w_1=
\frac{d^2w_1}{d\tau ^2}-w_1=0~.
\eqno(11)
$$
The particular solution $w_1=\cosh(\tau +\gamma)$ is usually called a 
zero mode. 
It is connected to the Riccati
solution through $u_1=\frac{1}{w_1}dw_1/d\tau$.
Next, changing the 
sign of the first derivative in ${\cal R}_1$ one calculates the outcome 
$\tilde{g}_f(\tau)$ using the same Riccati solution $u_1$, i.e.,
$-\frac{du_1}{d\tau}+u^2_{1}=2\tanh ^2(\tau +\gamma)-1:=\tilde{g}_f(\tau)$.
Thus, one considers a second Riccati equation
$$
{\cal R}_2: \quad -\frac{du}{d\tau}+u^2=2\tanh ^2(\tau+\gamma) -1~,
\eqno(12)
$$
of which we already know that it has $u_1$ as solution. 
The corresponding zero mode $w_2=1/\cosh (\tau +\gamma)$ fulfills
$$
\left(\frac{d}{d\tau}-u_1\right)
\left(\frac{d}{d\tau}+u_1\right)w_2=
\frac{d^2w_2}{d\tau^2}-\tilde{g}_f(\tau)w_2=0~.
\eqno(13)
$$

We are now interested in the general solution $u_{g2}$ 
of ${\cal R}_2$. 
To find it, one  employs the Bernoulli ansatz
$ u_{g2}(\tau)= u_1(\tau) - \frac{1}{f(\tau)}$, where $f(\tau)$ is an unknown
function. One obtains for the function $f(\tau)$ the
following Bernoulli equation
$$
 \frac{df(\tau)}{d\tau} + 2f(\tau)\, u_1(\tau) =1~,
\eqno(14)
$$
that has the solution
$$
 f(\tau)= \frac{{\cal I}_{0b}(\tau)+ \lambda}{w_1^{2}(\tau)}~,
\eqno(15)
$$
where $ {\cal I}_{01}(\tau)= \int _{0}^{\tau} \,
w_1^2(y)\, dy$,
and we consider $\lambda$ as a positive integration constant that is 
employed as a free parameter of the force field.

Thus, the general Riccati solution of ${\cal R}_2$ 
is a two-parameter function
$ u_{g2}(\tau; \gamma ,\lambda)$ of the following form
$$
 u_{g2}(\tau;\beta ;\lambda)= \frac{d}{d\tau}
\Big[ \ln \left(\frac{w_1(\tau)}{{\cal I}_{01}(\tau) +
\lambda}\right)\Big]=  \frac{d}{d\tau}
\Big[ \ln \left(\frac{\cosh(\tau+\gamma)}{\frac{1}{4}\sinh 2(\tau+\gamma)
+\frac{1}{2}(\tau+\gamma) +\lambda}\right)\Big]~,
\eqno(16)
$$
where in the last $\lambda$ we absorbed the lower limit of the integral.
The range of the $\lambda$ parameter is conditioned by ${\cal I}_{0}(\tau) +
\lambda \neq 0$ in order to avoid singularities. 
According to SIDT one can use $ u_{g2}$ to calculate a  
family of force functions 
$\tilde{g}_{\lambda}:=\frac{du_{g2}}{d\tau}+u_{g2}^2$. Thus
$$
\tilde{g}_{\lambda}=1-2\frac{d^2}{d\tau^2}\ln\left({\cal I}_{01}(\tau) 
+\lambda\right)~.
\eqno(17)
$$
This expression coincides with Eq.~(1) when passed to dimensional
quantities. Thus, there is a third Riccati equation, namely
$$
{\cal R}_3:\quad  \frac{du}{d\tau}+u^2=\tilde{g}_{\lambda}~,
\eqno(18)
$$
which is similar in form to ${\cal R}_1$ but 
possessing the solution $u_{g2}$. The corresponding linear equation is
$$
\left(\frac{d}{d\tau}+u_{g2}\right)
\left(\frac{d}{d\tau}-u_{g2}\right)w_3=
\frac{d^2w_3}{d\tau ^2}-\tilde{g}_{\lambda}w_3=0~,
\eqno(19)
$$
where
$$
w_3=\left(\frac{w_1(\tau)}{{\cal I}_{01}(\tau) +
\lambda}\right)~.
\eqno(20)
$$
In the limit $\lambda \rightarrow \infty$, Eq.~(18) goes 
into Eq.~(10) because $u_{g2}\rightarrow u_1$ and
$\tilde{g}_{\lambda}\rightarrow 1$. 
Thus, one can say that Eq.~(18) is the one-parameter SIDT
generalization of the Riccati equation (10). 

The problem of imposing initial conditions can be solved similarly to Davis. 
Since the solution $u_{g2}$ has two parameters, $\gamma$ and 
$\lambda$, we shall fix the phase parameter through the initial condition 
$u_{g2}(0)=u_0$ and the other parameter. 
This leads to the following cubic algebraic equation
$$
\tanh ^3 \gamma -u_0\tanh ^2 \gamma-\tanh \gamma =-(u_0+\frac{1}{\lambda})~.
\eqno(21)
$$
The three solutions of this equation can be obtained by Cardano`s formulas
and any of them can be used to 
fix $\gamma$. The real one is the most simple
$$
\gamma _1={\rm Arctanh}\frac{1}{3} 
\left(u_0-\frac{E^{1/3}-E^{-1/3}}{\lambda}\right)~,
\eqno(22)
$$
where $E=\left(\frac{B+\sqrt{4A^3+B^2}}{2}\right)^{1/3}$, 
$A=-3\lambda ^3-\lambda ^2 u_0$, and 
$B=27\lambda ^2+27 \lambda ^3u_0-9\lambda ^4u_0-2\lambda ^3u_0^3$.
The other two solutions are complex and being more complicated will not be 
reproduced here. 

Let us calculate the distance in the motion ${\cal R}_3$. It is given by
$$
s_{3}(\tau)=\int_{0}^{\tau}u_{g2}=
\ln \left(\frac{w_1(\tau)}{{\cal I}_{01}(\tau) +\lambda}\right)
\approx \ln \left(4e^{\tau}e^{-2\tau}\right)\rightarrow -Vt+\frac{2V^2}{g}\ln 2
\eqno(23)
$$
that shows that in the large-time asymptotics the motion is a vertical
rising of constant velocity $V$.

\bigskip

\section{Conclusion}

\bigskip

\noindent
One may say that if there exist planetary-like
astrophysical bodies with gravitational fields of the type given 
by Eq.~(1), then an anti-parachute effect might occur in their atmospheric
layer. It is hard to believe that nature can produce such fields.
For example, in the case of the common SIDT application to 
supersymmetric quantum mechanics, the scheme is used merely to extend
the class of exactly solvable potentials and in general is considered to be
a curiosity.
However, we point out the debate on the modifications of the 
gravitational fields in superconductors as a consequence of 
macroscopic quantum effects with applications to neutron stars.\cite{Umm} 

\noindent
Finally, we would like to make the following comment.
SIDT introduces singularities in all the transformed quantities.
The parameter $\lambda$ gives 
the measure of the time interval where the singularity is located in the past
($t<0$). A small $\lambda$ means a singularity close to the moment 
of parachute jumps and the time dependence is significant. 
On the other hand, a large $\lambda$ leads to very small variation in time of
the planetary gravitational field that can be thought of as a cosmological 
evolution implying a singularity very far in the past.

\bigskip
\bigskip
\bigskip
\bigskip


\noindent
The authors are much indebted to Dr. M. Nowakowski for discussions and 
active criticism.

\bigskip
\noindent

\bigskip

{\bf Appendix}

\bigskip

\noindent
In general, for a Riccati equation of the type 
$$
 \frac{dy}{dx}=\alpha y^2+\beta~, \qquad  \alpha \, , \beta = {\rm const.}
\eqno(A1)
$$
the solutions can be written as follows

\bigskip

(i) ${\rm for} \quad  \frac{\alpha}{\beta}>0$
$$
y=\sqrt{\frac{\beta}{\alpha}}\tan\left
(\sqrt{\frac{\alpha}{\beta}}\beta x+\phi _1\right),
\qquad {\rm or} \qquad 
y=\sqrt{\frac{\beta}{\alpha}}\cot
\left(-\sqrt{\frac{\alpha}{\beta}}\beta x+\phi _2\right)
\eqno(A2)
$$

\bigskip

(ii) ${\rm for} \quad \frac{\alpha}{\beta}< 0$
$$
y=\sqrt{-\frac{\beta}{\alpha}}\tanh
\left(\sqrt{-\frac{\alpha}{\beta}}\beta x+
\varphi _1\right),\quad {\rm or} \quad 
y=\sqrt{-\frac{\beta}{\alpha}}\coth\left(\sqrt{-\frac{\alpha}{\beta}}\beta x+
\varphi _2\right)~.
\eqno(A3)
$$
 
Comparing the Newtonian falling case with Eq.~(A1) one gets
$$
\alpha =-\epsilon, \qquad \beta =g \qquad \Longrightarrow \quad
\frac{\alpha}{\beta}=
\frac{-\epsilon}{g}<0,
\eqno(A4)
$$
and therefore the solution is given by
$$
v(t)=\sqrt{\frac{g}{\epsilon}}\tanh\left(\sqrt{\epsilon g}t+
\varphi _1\right)~,
\eqno(A5)
$$
which is Davis' result. The falling velocity $v(t)$ presents a substantial
time dependence - acceleration stage - only for a short period and rapidly 
turns into the limiting constant velocity $V=\sqrt{\frac{g}{\epsilon}}$.  

\newpage


\vskip 2ex
\centerline{
\epsfxsize=180pt
\epsfbox{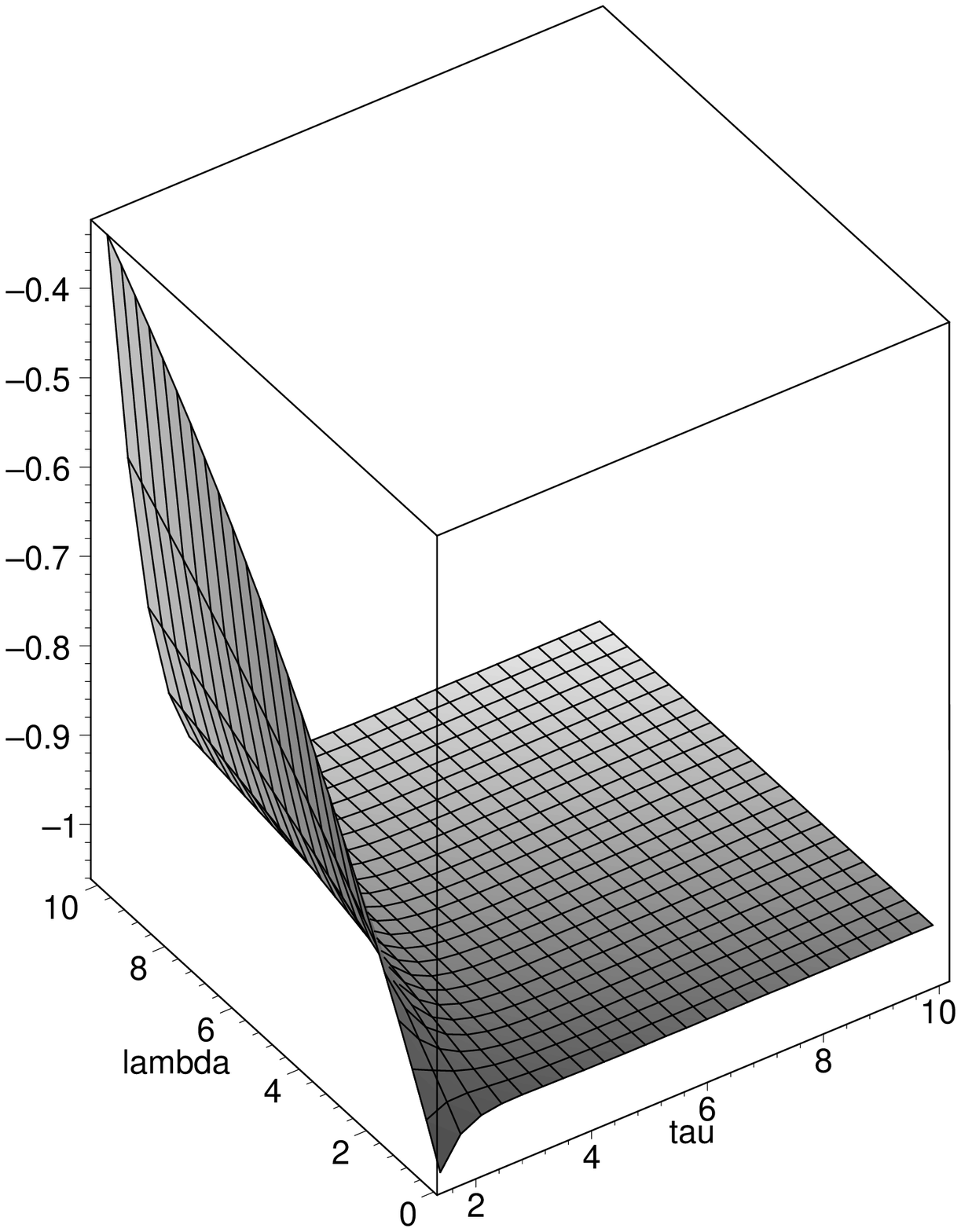}}
\vskip 4ex
\begin{center}
{\small{Fig. 1:}$\quad$
The velocity $u_{g2}$ for $\gamma _1=1$.}
\end{center}

\vskip 2ex
\centerline{
\epsfxsize=180pt
\epsfbox{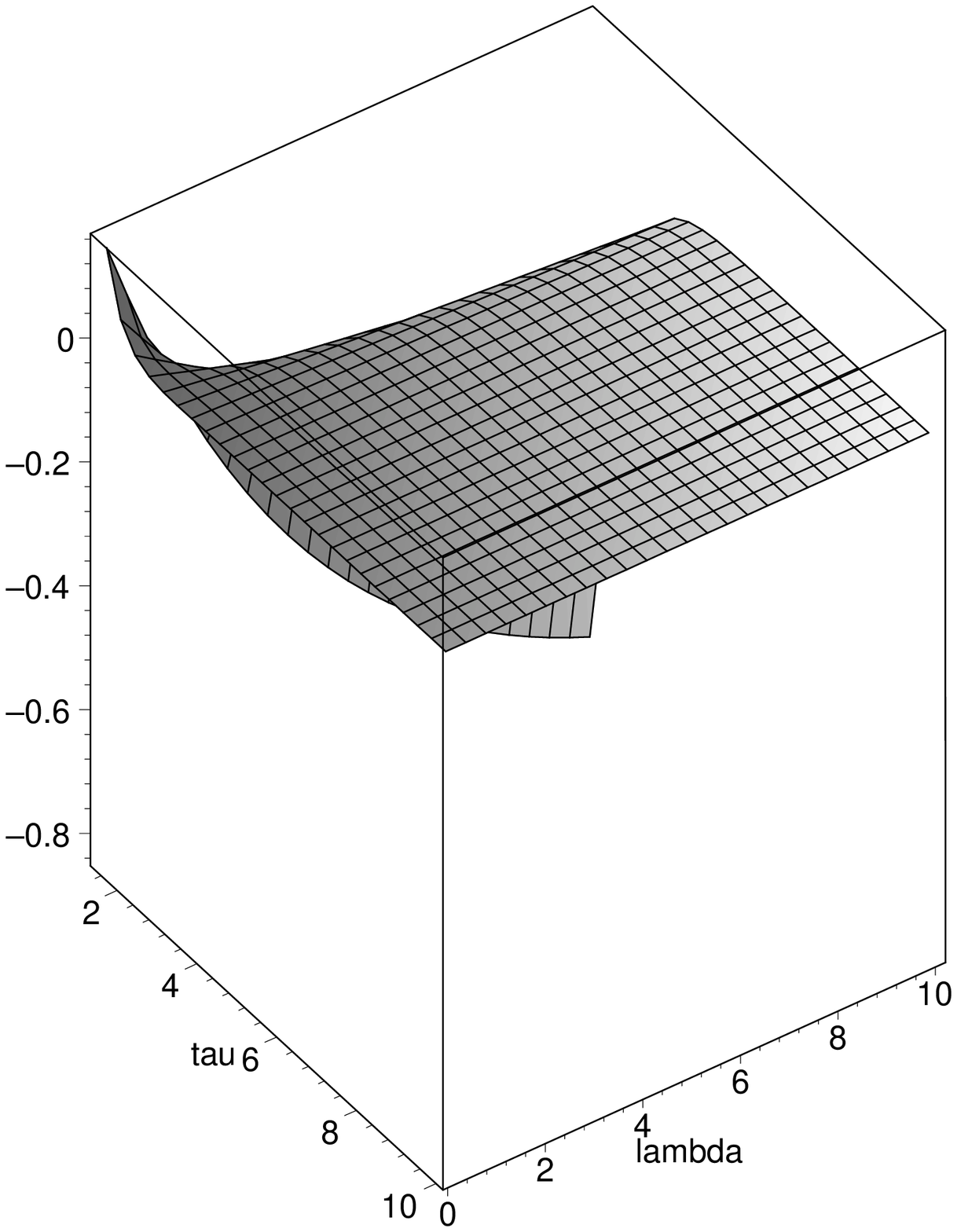}}
\vskip 4ex
\begin{center}
{\small{Fig. 2:}$\quad$
The derivative of $u_{g2}$ (the acceleration) for the same $\gamma _1$.
}
\end{center}

\vskip 2ex
\centerline{
\epsfxsize=180pt
\epsfbox{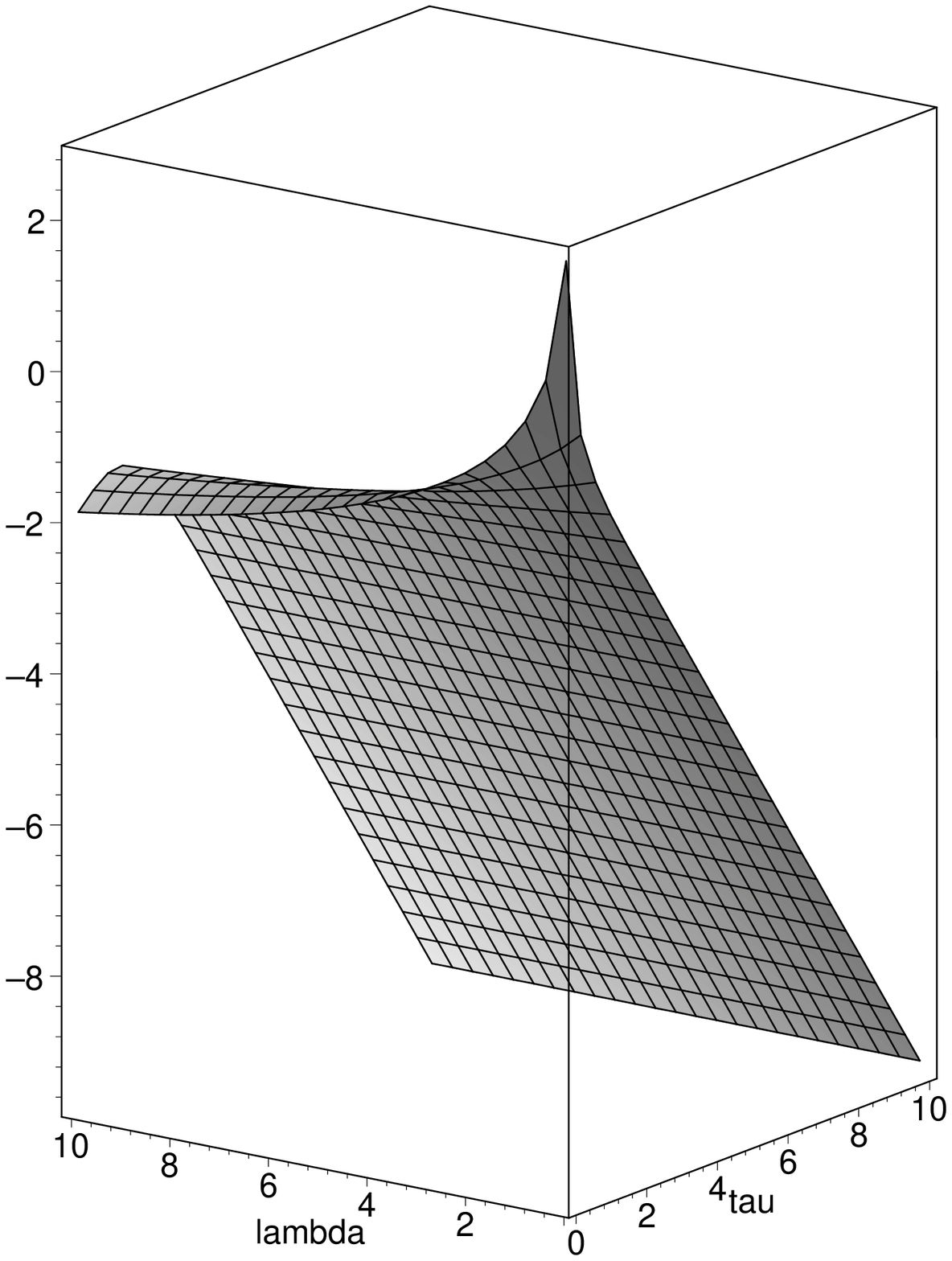}}
\vskip 4ex
\begin{center}
{\small{Fig. 3:}$\quad$
The rising distance $s_3$ for the same $\gamma _1$.
}
\end{center}

\vskip 2ex
\centerline{
\epsfxsize=180pt
\epsfbox{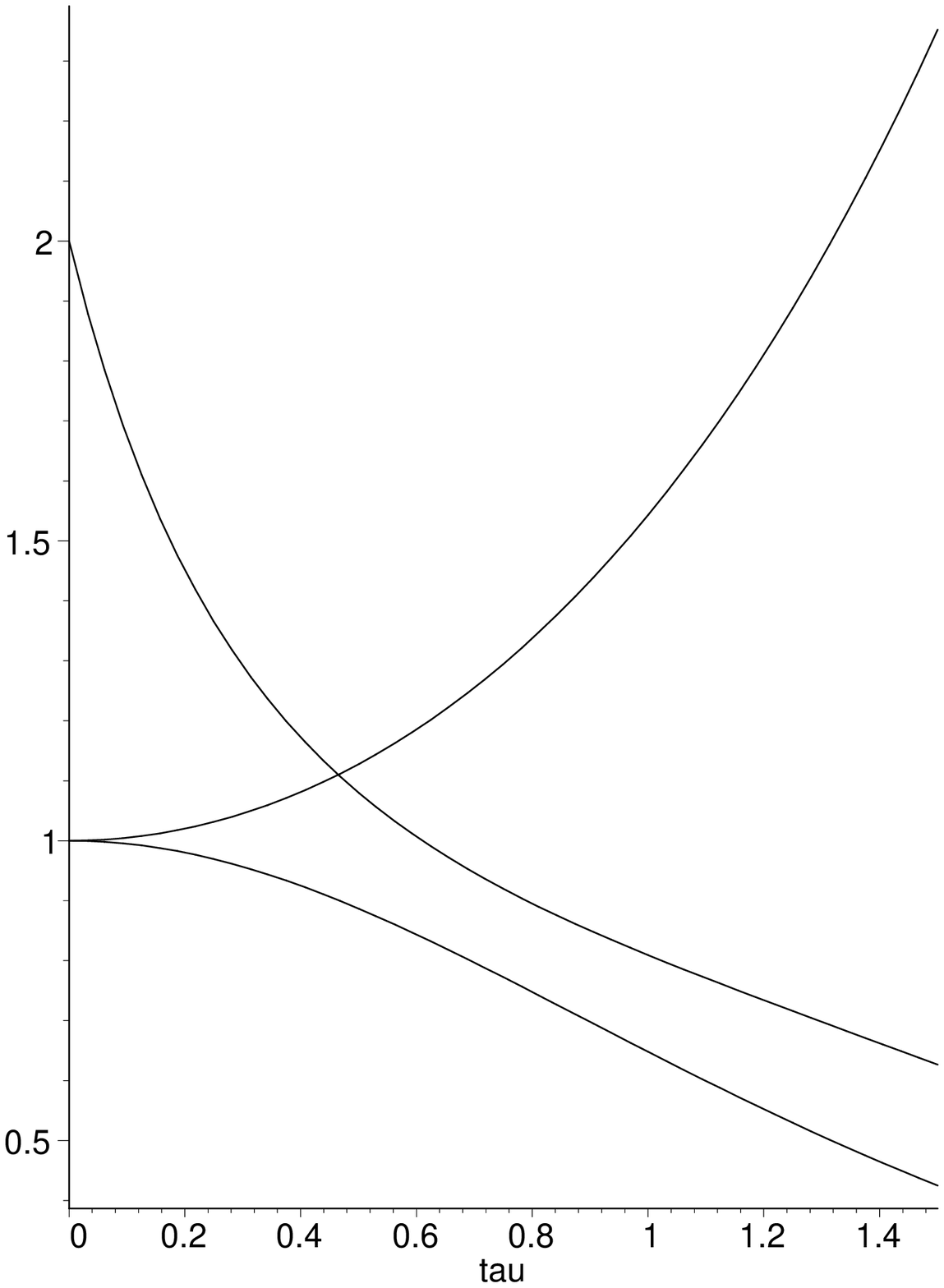}}
\vskip 4ex
\begin{center}
{\small{Fig. 4}$\quad$
The zero modes $w_1$, $w_2$, and $w_3(\lambda =0.5)$.}
\end{center}

\vskip 2ex
\centerline{
\epsfxsize=180pt
\epsfbox{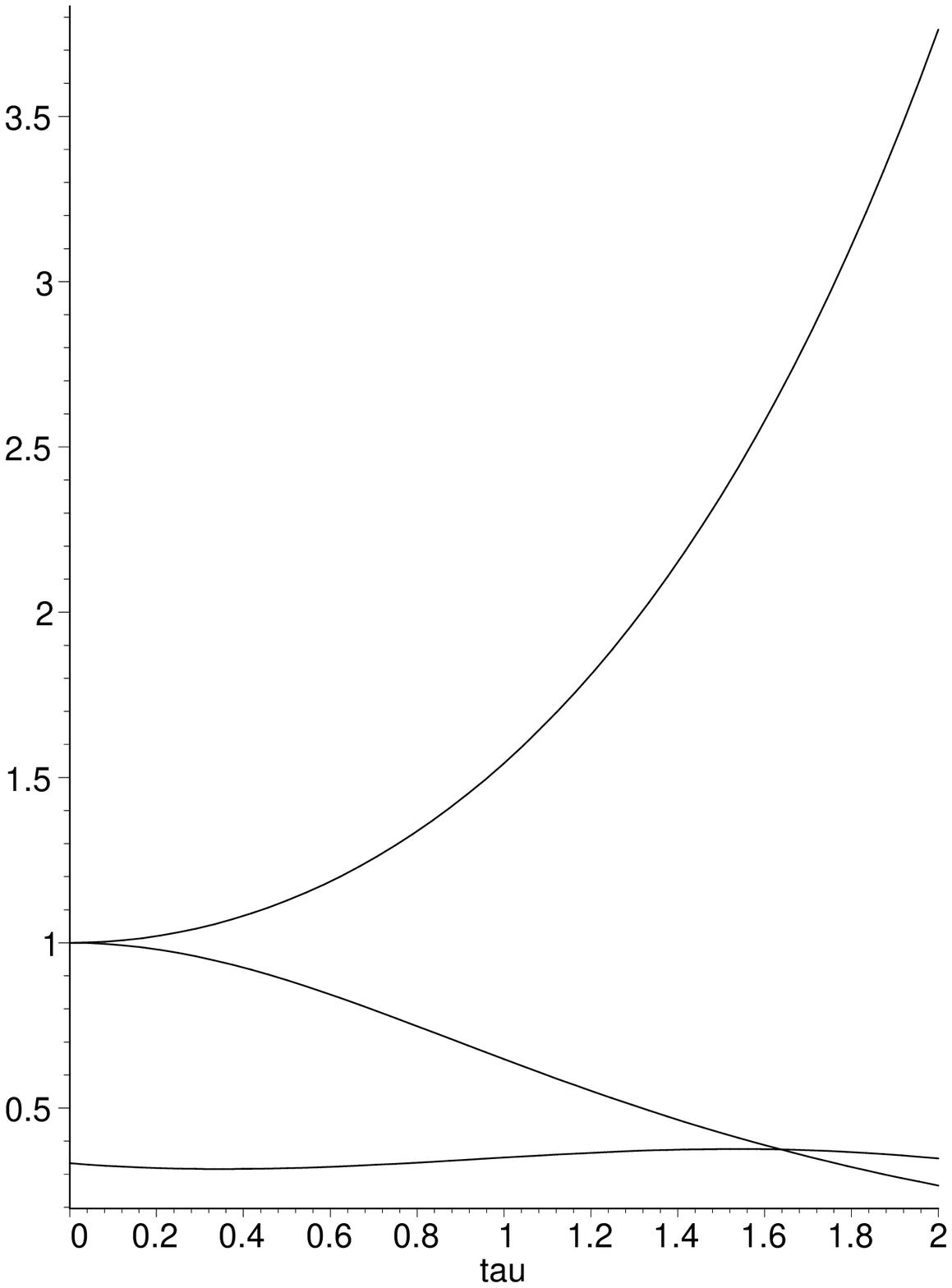}}
\vskip 4ex
\begin{center}
{\small{Fig. 5}$\quad$
The zero modes $w_1$, $w_2$, and $w_3(\lambda =3)$.}
\end{center}

\vskip 2ex
\centerline{
\epsfxsize=180pt
\epsfbox{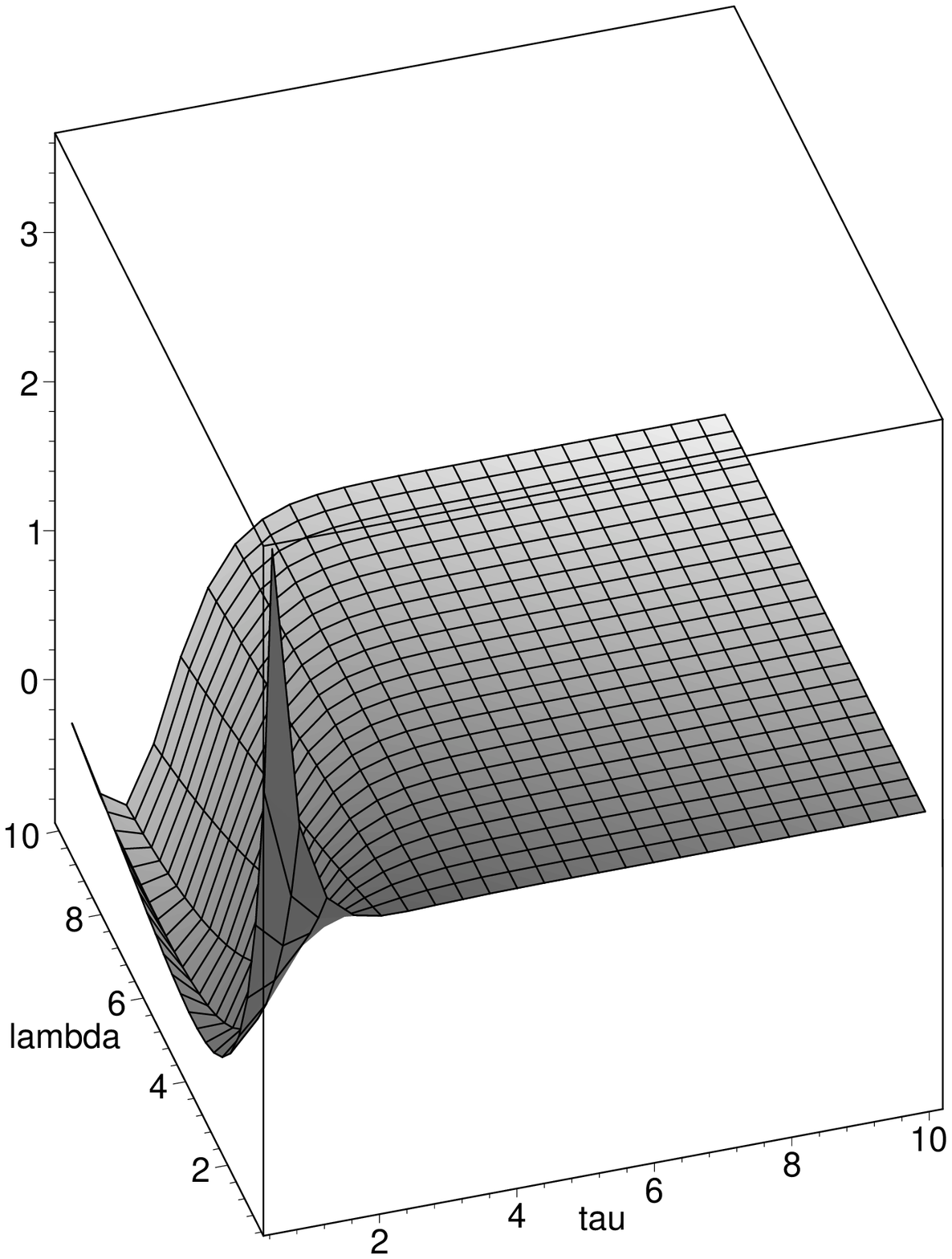}}
\vskip 4ex
\begin{center}
{\small{Fig. 6}$\quad$
The normalized force $\tilde{g}_{\lambda}$ for $\gamma _1=1$.}
\end{center}




\end{document}